\documentclass[useAMS,usenatbib]{mn2e}
\usepackage{graphicx}

\newcommand{\kms}{~km~s$^{-1}$ }
\newcommand{\apj}{ApJ}
\newcommand{\xspec}{{\small XSPEC~}}

\newcommand{\snr}{~SNR~1987A~}
\newcommand{\snrE}{~SNR~1987A}

\title[Evolution of the {\it Chandra} CCD Spectra of \snr]
{Evolution of the {\it Chandra} CCD Spectra of \snr: Probing the 
Reflected-Shock Picture}
\author[S.A.Zhekov et al.]{Svetozar A. Zhekov$^{1,4}$
\thanks{E-mail: zhekovs@colorado.edu (SAZ); park@astro.psu.edu (SP);
richard.mccray@colorado.edu (RMcC); judith.racusin@nasa.gov (JLR);
burrows@astro.psu.edu (DNB)}, 
Sangwook Park$^2$, Richard McCray$^1$, Judith L.  Racusin$^3$ and 
\newauthor David N. Burrows$^2$\\
$^1$JILA, University of Colorado, Boulder, CO
80309-0440, USA\\
$^2$Department of Astronomy and Astrophysics,
Pennsylvania State University, 525 Davey Laboratory, University
Park, PA 16802, USA\\
$^3$NASA GSFC, 8800 Greenbelt Rd., Code 661, Greenbelt, MD 20771, 
USA\\
$^4$On leave from Space Research Institute, Sofia, Bulgaria
}
\date{}

\pagerange{\pageref{firstpage}--\pageref{lastpage}} \pubyear{2002}

\begin{document}

\maketitle

\label{firstpage}

\begin{abstract}
We continue to explore the validity of the reflected shock structure
(RSS) picture in \snr that was proposed in our previous analyses of 
the X-ray emission from this object. We used an improved version of 
our RSS model in a global analysis of 14 CCD spectra from the 
monitoring program with {\it Chandra}. 
In the framework of the RSS picture,
we are able to match both the expansion velocity curve
deduced from the analysis of the X-ray images  and
light curve.
Using a simplified analysis, we also show that the X-rays and the
non-thermal radio emission may originate from the same shock structure
(the blast wave).
We believe that using the RSS model in the analysis of grating data
from the {\it Chandra} monitoring program  of
\snr that cover a long enough time interval, 
will allow us to build 
a more realistic physical picture and model of \snrE.
\end{abstract}


\section{Introduction}
After its reappearance in radio (\citealt{turtle_90};
\citealt{stav_92}, 1993) and in X-rays (\citealt{buer_94};
\citealt{gor_94}; \citealt{has_96}) at about 1000-1200 days after 
the explosion (DAE), SN 1987A entered into the phase 
of supernova remnant (SNR) formation \citep{dick_07}. Observations of 
this phenomenon from its onset give us a chance to study 
in detail the physics of strong shocks as well as to investigate the 
distribution of the circumstellar matter (CSM) around the exploded 
star. The latter may help us solve the mystery of the origin of 
the triple-ring system observed in the optical (\citealt{wampler_90};
\citealt{jak_91}; \citealt{crotts_91}; \citealt{burr_95}).
With this respect, X-ray observations are very important since 
they manifest most of the energetics at present and they
provide us with direct information about the underlying physics of
this evolving object.  

The richness and variety of processes involved in
the birth of \snr are manifested by observations across the entire 
electromagnetic spectrum. Radio observations
(\citealt{stav_92}, 1993, 2007; \citealt{ball_01}; 
\citealt{man_02}, 2005; \citealt{gaensler_97}, 2007; \citealt{ng_08}; 
\citealt{potter_09}; \citealt{zana_10}) 
tell us about the non-thermal (synchrotron)
emission mechanisms and relativistic particle acceleration in strong 
shocks. Near-infrared observations tell us 
about the distribution of dust in the CSM and the physics of 
its interaction with the hot plasma (\citealt{bou_04}, 2006; 
\citealt{dwek_08}). Optical and UV observations (\citealt{son_98}; 
\citealt{law_00}; \citealt{pun_02}; \citealt{suger_02};
\citealt{gron_06}, 2008)
tell us about the physics of slow shocks and the distribution of dense
clumps of gas.

The early evolution of \snr is strongly influenced by the distribution
of circumstellar gas, which is concentrated in its triple-ring system.
Although the origin of the ring system is not well understood, the
colliding wind scenario that assumes a spherically-symmetric blue
supergiant (BSG) wind interacting with an earlier emitted asymmetric 
red supergiant
(RSG) wind seems the physically most reasonable explanation
(\citealt{luo_91}; \citealt{wang_92}; \citealt{morris_07}). 
It has been explored in detail 
by \citet{bl_lu_93} who demonstrated that such a model is
capable of accounting for the observed CSM characteristics,
provided the RSG wind was highly asymmetric with most of its mass
confined near the equator. This picture needed an additional 
ingredient in order to explain the reappearance of \snr in radio 
and X-rays. \citet{che_dwa_95} proposed that an extended 
HII region was formed inside the inner equatorial ring that was
naturally produced by the photoionization of the shocked RSG wind 
(the inner ring) by the central progenitor BSG star. Based on this,
Borkowski et al. (1997a, b) studied the complex hydrodynamics of the
interaction of the SN ejecta with the CSM in vicinity of the inner
ring that also predicts a continuous brightening of \snr in UV and
X-rays.

Since its reappearance, \snr has been brightening continuously 
in X-rays with an accelerating rate in 
the past few years (\citealt{sang_05}, 2006, 2007; \citealt{judy_09}). 
The first spectrum of \snr with high spectral resolution was obtained
with the high energy transmission grating on {\it Chandra} by
\citet{eli_02}.
The subsequent  brightening has enabled dispersed spectra with good
photon statistics to be obtained with both
{\it Chandra} (\citealt{zhek_05}, 2006; \citealt{dew_08}; 
\citealt{zhek_09}) and {\it XMM-Newton} (\citealt{haberl_06}; 
\citealt{heng_08}). With the superb spatial and spectral 
resolution of the {\it Chandra} observatory, the first direct evidence 
was obtained that the distribution of the X-ray emitting plasma in 
\snr is not spherically-symmetric but it is `flat', {\it i.e.,} it is 
concentrated in the vicinity of the equatorial ring 
(\citealt{zhek_05}, 2006; \citealt{dew_08}; \citealt{zhek_09}).
These authors also showed that 
the bulk gas velocities deduced from the X-ray spectral lines are 
too low to account for the plasma temperatures inferred from the
spectral fit, and showed that a simplified reflected-shock structure
(RSS) model could match the observed spectra successfully.

Superb spatial resolution of {\it Chandra} allowed \citet{burr_00} 
to obtain the first  image of \snr that showed a ring-like structure 
in X-rays. Subsequently, a monitoring program was established with 
{\it Chandra} to track the evolution of the X-ray images and CCD 
spectra of \snrE. In addition to  the continuing brightening in 
X-rays, the spectral data showed a gradual decrease of the plasma 
temperature  behind the fast shocks. From the imaging data, we 
measured the expansion velocity of the remnant (\citealt{sang_02}, 
2004, 2007).  After applying a physically more realistic method 
to analyze the X-ray images, \citet{judy_09} obtained a 
more accurate expansion velocity curve. It showed clearly that 
the velocity started to decelerate after $\sim 6000$~DAE, about the
same time that the X-ray light curve turned up \citep{sang_05}.

Here we analyze the evolution of the X-ray spectrum of \snr as
observed with the pulse-height spectra obtained with the ACIS CCD
camera during the monitoring program.  One of the main goals of our
study is to test the reflected-shock structure picture
that emerged from analysis of the
{\it Chandra} grating data. 
Since the pulse-height spectra have much better time coverage 
than the grating data, they can test 
whether is possible to match the CCD spectra
and the expansion velocity curve in the framework of the RSS picture.
In section \S~\ref{sec:rss_picture} we give a qualitative description
of the RSS picture for \snrE. In section \S~\ref{sec:rss_model} we
present details about the global RSS model. We describe the evolution
of the CCD spectra and results from the spectral fits in sections 
\S~\ref{sec:data} and \S~\ref{sec:spectral_fits}, respectively.
In section \S~\ref{sec:discussion} we discuss the results, how they
might be related to the non-thermal radio emission from \snr and the
consistency of the global RSS picture. 
We present our conclusions in section \S~\ref{sec:conclusions}.

\section{Physical picture and motivation}
\label{sec:rss_picture}
The physical picture that emerges from the analysis of \snr so far 
is the following. The circumstellar matter around the inner 
equatorial ring likely consists of at least two components: 
(1) a smooth gas component which represents the so called HII region 
(resulting from evaporation of the inner ring or being the hot bubble 
in the interacting stellar winds scenario); 
and (2) much denser clumps distributed within the smooth component.
Although the origin of the dense clumps is unclear, it seems
realistic to expect that they will form as a result from various
dynamical instabilities. As \citet{bl_lu_93} point out, in the
interacting stellar winds scenario these
instabilities are expected to produce a very clumpy and nonuniform
shell. Thus, during the shell evolution 
dense clumps penetrate into the smooth component and 
form protrusions of various shapes.

Interaction of the SN ejecta with the smooth component gives rise to
the standard two-slab structure bounded by a forward shock (blast wave
compressing the smooth CSM component) and a reverse shock (compressing
the ejecta material). The slabs are separated by a contact
discontinuity. 

When the blast wave overtakes a clump, it is `lit up', namely, a
transmitted shock starts to heat up the clump gas. At the same time,
a reflected shock
is sent back that shocks the gas behind the blast wave again,
further increasing  its
temperature and  decreasing its bulk velocity. Such a
reflected-shock structure  was invoked to explain the relatively
narrow spectral lines detected by the {\it Chandra} grating observations 
of \snr and the corresponding simplified model was successful in 
describing these results. 
It should be kept in mind that the actual hydrodynamics of such an
interaction is very complex: the clumps may have various shapes;
the blast wave may wrap up around a clump and result in a more
complicated shock structure; the reflected shocks from adjacent clumps
may additionally interact to create a
`turbulent' system of gas shocked multiple times. 
A proper description of such interaction would require
3D hydrodynamic calculations. Thus, the RSS model
might be considered as a very simple representation of the actual
physical picture.

On the other hand, it does not seem feasible to attempt a complete
match of the \snr observables (X-rays spectra, light curve etc.) by a
detailed hydrodynamic model mostly because there are too many initial
(and boundary) conditions to explore. Instead, we think it would be
instructive to use a simplified model that bears the basic physics
of this complex interaction and by using such a tool we hope to
deduce valuable information about the CSM in the vicinity of the inner
ring of \snrE. Results from such an analysis of the available (and
future) observations can guide us towards building a more realistic 
picture (and model) of this exciting phenomenon: the formation and 
development of \snrE.

\section{Global RSS Model}
\label{sec:rss_model}
\citet{zhek_09} proposed 
a reflected-shock structure model to interpret
the {\it Chandra} grating data for \snr taken in 2004-2007.
It bears the
basic physics of interaction of a shock with a dense clump which is a
cornerstone in the physical picture described above.
Since the CCD monitoring observations provide more extensive time
coverage of the \snr evolution, they can be used to test details of
this physical picture. To do so, we adopt the following improvements 
of the RSS model:

\begin{itemize}
\item
At every moment (observation) the entire (integrated) X-ray
characteristics of \snr can be represented with a RSS structure (blast
wave, transmitted and reflected shocks), i.e., the
blast wave is impinging on some average clump that is representative
of the clumps at that radius from the center of the ring
(the mean of the clumps distribution).

\item
The emission measure (EM) of the blast wave is not a free parameter from
one observation to another but it follows the density profile of the
smooth component (EM~$\propto n^2 r^3$, where $n$ is the density of the
smooth component and $r$ is the radius of the \snr at that moment).

\item
The \snr radius is calculated from 1D analytical solution that takes
into account the density profile of the smooth component.

\item
The electron and ion temperatures are not equal behind the shocks
(but see below).

\end{itemize}

\begin{figure}
\centering\includegraphics[width=3.in,height=2.in]{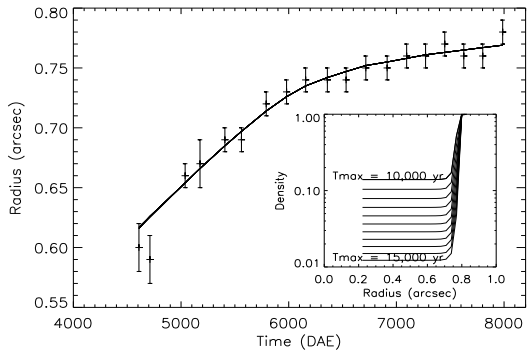}
\caption{
The analytical solution fits to the
radial-expansion curve from \citet{judy_09}.
The solid line is an overlay of a few solutions having $t_{max}$ from
10,000 to 15,000. The corresponding density profiles of the HII region
are shown in the inset figure.
}
\label{fig:cheva}
\end{figure}

\subsection{ Analytical solution}
\label{subsec:anal_sol}
We anticipate that the density profile of the smooth component will 
be increasing steeply towards the inner ring radius as measured 
from the {\it Hubble Space Telescope (HST)} images. 
Many choices of such a function are possible and the model chosen 
here is:

\begin{equation}
 \rho = \rho_0 + (\rho_{Ring} - \rho_0) \
tanh \left(D\frac{r - R_0}{R_{Ring} - r}\right)^s 
\label{eq:rho}
\end{equation}
where $R_{Ring} = 0.83$~arcsec \citep{suger_02} is the radius of the 
equatorial ring; 
$R_0$ is the inner radius of the HII region;
$\rho_{Ring}$ and $\rho_0$ are the densities respectively at
these radii; $D$ and $s$ control the shape of the density profile.
We assume that the outer supernova envelope has a
density profile, $\rho \propto t^{-3}(r/t)^{-n}$, 
where $n = 9$ in the case of \snr (e.g.,
\citealt{east_89}; \citealt{suz_93}; \citealt{bor_97a},b).
\citet{cheva_82} derived the solution for
the structure that results from interaction of such a density profile
with the circumstellar medium.
Guided by his approach (i.e., using dimensional analysis), 
we derive an analytical
solution for the radius, $r$, of the structure (the contact
discontinuity). In the case of the CSM density profile 
given by eq.~(\ref{eq:rho}) we find:

\begin{eqnarray}
 t = t_{max} \left[\tilde{\rho_0} +(1 - \tilde{\rho_0})
tanh \left(D\frac{r - R_0}{R_{Ring} -
r}\right)^s\right]^{\frac{1}{n-3}} \times \nonumber \\
\left(\frac{r}{R_{Ring}}\right)^{\frac{n}{n-3}}
\label{eq:anal}
\end{eqnarray}
where $\tilde{\rho_0} = \rho_0/\rho_{Ring}$ and $t_{max}$ is the time
when the radius of the blast wave equals that of the ring.

We found that this expression for $r$ is in satisfactory agreement
with the result of a 1D hydrodynamic simulations for 
the case of stellar ejecta with $n = 9$ interacting with CSM that has
density profile as given in eq.~(\ref{eq:rho}). 
This makes us
confident in using the analytical solution in our model.

Figure ~\ref{fig:cheva} shows that
equation~(\ref{eq:anal}) gives a good fit to the observed
radial-expansion curve \citep{judy_09}.
Note that there is not a unique set of fit
parameters ($t_{max}$, $\tilde{\rho_0}, D$ and $s$) that represents
the `best' fit.  Namely, parameters $t_{max}$  and $\tilde{\rho_0}$ 
cross talk and this is illustrated in Fig.~\ref{fig:cheva} where the
solid curve is in fact an overlay of a few solutions having 
$t_{max}$ in the range from 10,000 to 15,000 DAE (the corresponding
density profiles are shown in the inset figure).
This behavior is well understood since the fit to the
radial-expansion curve is determined by the relative change of the 
density of the smooth component between the observations. 
It will become possible to discriminate among 
different cases as the blast wave progresses
further into the equatorial ring.

\subsection{\xspec model}
\label{subsec:xspec_model}
Based on the analytical solution, we extended the RSS model for
\xspec\footnote{Version 11.3.2 of the \xspec analysis
package \citep{Arnaud96} is used in this study.} 
by  \citet{zhek_09}. The new RSS model 
tracks the radius of the blast wave that
is inferred from the observations \citep{judy_09}. The emission 
measure behind this shock is scaled correspondingly 
(EM~$\propto n^2 r^3$) and its ionization time ($n_e t$) follows the
density profile of the preshock gas (in fact, its integral over
time for a particular observation). 
Also, the
blast wave velocity is estimated for each observation. Bearing in mind
that the blast wave velocity has decreased recently from a few
thousands \kms \citep{judy_09} and the fact that the 
electron temperature of the X-ray emitting plasma was not greater than
3 keV \citep{sang_06}, we conclude that the electron and ion
temperatures in the plasma behind the
blast wave have not completely equilibrated
(the mean plasma temperature 
kT$=1.4\times[V_{shock}/1000~\mathrm{km~s^{-1}}]^2$~keV for the 
\snr abundances).
We recall that in the original RSS model, \citet{zhek_09} assumed a
complete temperature equilibration (1T approximation) and took into
account the non-equilibrium ionization (NEI) effects explicitly. 
Here we adopted the same approximation which can be justified by the 
following considerations.

First, a direct comparison  between the
X-ray spectra from a plane-parallel shock (PPS) model with different
electron and ion temperatures (2T approximation; {\it vnpshock} in 
\xspec) and from a 1T PPS model ({\it vpshock} in \xspec) shows that 
the former model can be matched well by the latter having a postshock 
temperature averaged over
the shock ionization age (see also \citealt{bor_01}). This simply
indicates that the NEI effects are more pronounced than the 
effects due to the electron temperature evolution downstream from the 
shock front. Second, we constructed a RSS model that correctly treats 
the temperature
equilibration behind the shock (similarly as in {\it vnpshock} in 
\xspec). But we found that its application to the entire set of 
monitoring observations is very unstable because the 
parameter space is too complicated.

Thus, the new RSS model correctly estimates the mean plasma temperature
behind the blast wave. That result is then used to calculate the mean 
plasma
temperature behind the transmitted shock (as well as behind the 
reflected shock: see Appendix B in \citealt{zhek_09}).  
The spectral fits assume an average electron temperature 
($\bar{T_e}$) typical for either of the shocks. From these
spectral fits we derive a parameter which is the average 
$\beta = \bar{T_e}/T$ ($T$ is the mean plasma temperature) for 
each shock and its value sets an upper limit to the actual 
$\beta$ in the shock front. We note that the reflected shock is 
assumed to have the same average $\beta$ as the blast wave.
As discussed by \citet{zhek_09}, the plasma behind the reflected shock 
has quite complicated heating history. It is so since the reflected
shock interacts with the plasma behind the blast wave that has already
evolved for some time. As a result, in the region behind the reflected
shock we have a mixture of hot plasma
with different heating histories, resulting in a distribution of 
temperatures and ionization time scales.
To simplify the model 
calculations and also due to the technical problems mentioned above, 
we adopted an {\it average} $\beta$ parameter for the reflected shock 
that is equal to that in the blast wave.

\section{CCD Spectra}
\label{sec:data}
The CCD spectra from the monitoring program of \snr are exactly the
same (source and background spectra, redistribution and ancillary 
response functions) as in \citet{sang_06} and \citet{sang_07}. 
Details of data reduction and spectral extraction are found therein.
In this work we analyze 14 data sets,
with {\it Chandra} ObsIds as follows:
122, 1967, 1044, 2831, 2832, 3829, 3830, 4614, 4615, 5579 and 6178, 
5580 and 6345, 6668, 6669, 7636 (two of the observations were split
into two subsets).

\begin{figure*}
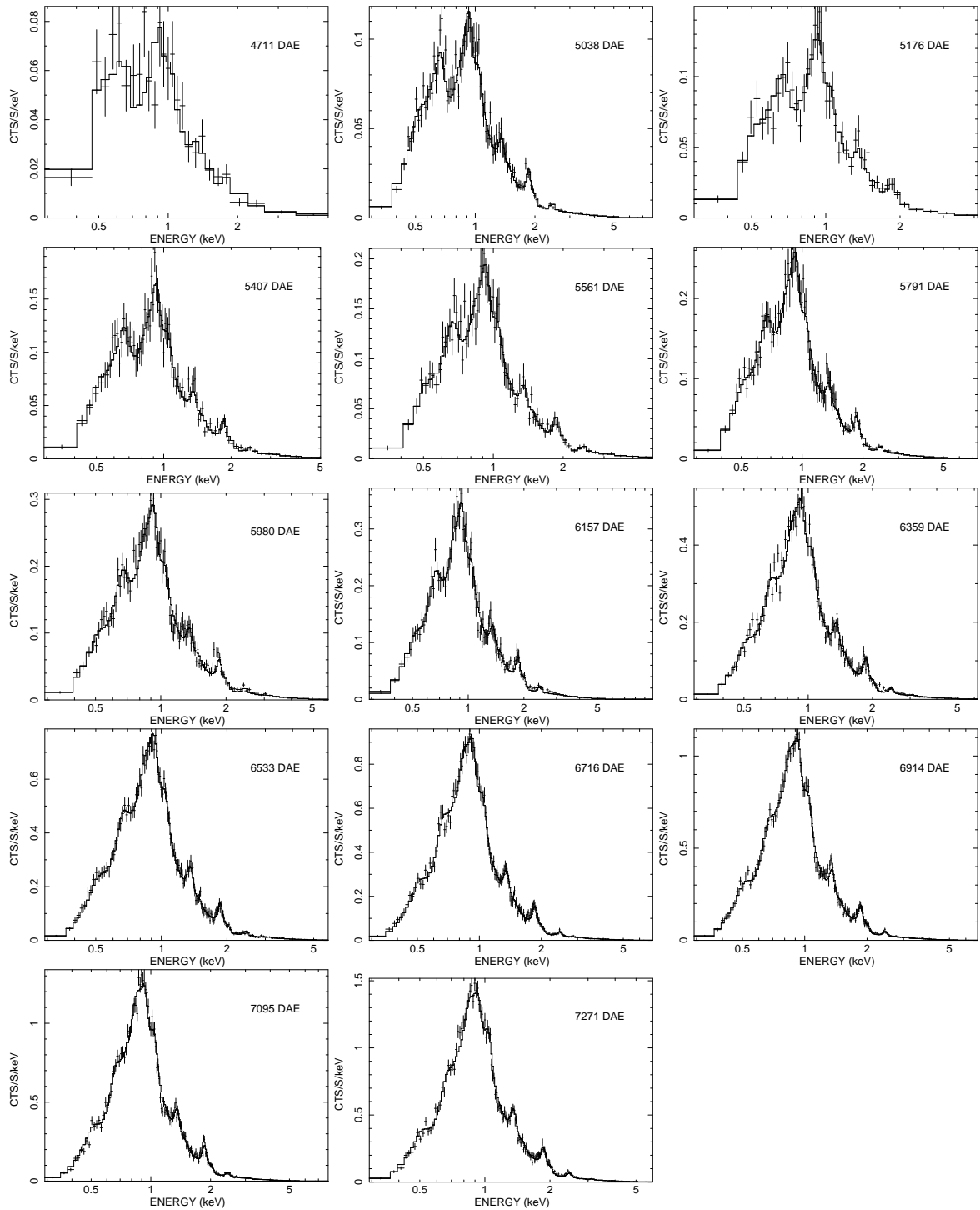

 \centering\includegraphics[width=1.5in,height=2.in,angle=-90]{fig2a.eps}
 \centering\includegraphics[width=1.5in,height=2.in,angle=-90]{fig2b.eps}
 \centering\includegraphics[width=1.5in,height=2.in,angle=-90]{fig2c.eps}

 \centering\includegraphics[width=1.5in,height=2.in,angle=-90]{fig2d.eps}
 \centering\includegraphics[width=1.5in,height=2.in,angle=-90]{fig2e.eps}
 \centering\includegraphics[width=1.5in,height=2.in,angle=-90]{fig2f.eps}
 \centering\includegraphics[width=1.5in,height=2.in,angle=-90]{fig2g.eps}
 \centering\includegraphics[width=1.5in,height=2.in,angle=-90]{fig2h.eps}
 \centering\includegraphics[width=1.5in,height=2.in,angle=-90]{fig2i.eps}
 \centering\includegraphics[width=1.5in,height=2.in,angle=-90]{fig2j.eps}
 \centering\includegraphics[width=1.5in,height=2.in,angle=-90]{fig2k.eps}
 \centering\includegraphics[width=1.5in,height=2.in,angle=-90]{fig2l.eps}
 \centering\includegraphics[width=1.5in,height=2.in,angle=-90]{fig2m.eps}
 \centering\includegraphics[width=1.5in,height=2.in,angle=-90]{fig2n.eps}
 \centering\includegraphics[width=1.5in,height=2.in,angle=-90]{fig2o.eps}
\caption{
The background-subtracted spectra of \snr overlaid with the model
fits for the case with $t_{max} = 13,000$ DAE (see
\S~\ref{sec:spectral_fits}).
Each panel is marked by the
days after explosion of the corresponding observation.
All the spectra are re-binned to have a minimum of 50 counts per bin
but the observations at 4711 DAE (20 cts/bin) and at 5178 DAE
(30 cts/bin).
}
\label{fig:spectra}
\end{figure*}

\section{Spectral Fits}
\label{sec:spectral_fits}
As in previous studies of the {\it Chandra} spectra of\snrE, we assumed 
that the X-ray absorption and abundances of the hot plasma do not 
change with time.  Thus, all spectra shared the value for $N_H$ and 
had the same abundances.
The procedures for fitting abundances are the same as those 
described in \citet{eli_02}. Namely, we only varied
the abundances of elements having strong emission lines in the observed 
(0.5 - 4 keV) energy range: N, O, Ne, Mg, Si, S and Fe. 

\begin{table}
\caption{The RSS Model Results (CCD Spectra)}
\label{tab:fit}
\begin{tabular}{lll}
\hline
$\chi^2$/dof  & 1706/1473 &  \\
N$_H$(10$^{21}$ cm$^{-2}$) & $2.23\pm0.05$  &  \\
H    & 1  & 1 \\
He        & 2.57  &  2.57 \\
C~         & 0.09  & 0.09  \\
N~(1.63)   &  $0.32\pm0.04$ &  0.56 [0.50 - 0.65]  \\
O~(0.18)   &  $0.07\pm0.01$ &  0.081 [0.074 - 0.092]  \\
Ne(0.29)   &  $0.22\pm0.01$ &  0.29 [0.27 - 0.31]  \\
Mg(0.32)   &  $0.17\pm0.01$ &  0.28 [0.26 - 0.29]  \\
Si(0.31)   &  $0.26\pm0.01$ &  0.33 [0.32 - 0.35]  \\
S~(0.36)   &  $0.36\pm0.02$ &  0.30 [0.24 - 0.36]  \\
Ar   & 0.537  & 0.537 \\
Ca   & 0.339  & 0.339 \\
Fe(0.22)   &  $0.10\pm0.01$ &  0.19 [0.19 - 0.21]  \\
Ni   & 0.618  & 0.618 \\
\hline
\end{tabular}

\vspace{0.5cm}
Note --  Results from the {\it simultaneous} fit to  14 CCD spectra
of \snrE.
The uncertainties are $1\sigma$ errors from the fit.
All abundances are expressed as ratios to their solar values
\citep{an_89}.
For comparison, the inner-ring abundances of He, C, N, and O
\citep{lf_96}  and
those of Ne, Mg, Si, S and Fe typical for LMC SNRs \citep{hu_98}
are given in the first column in parentheses.
The Ar, Ca and Ni abundances are representative for LMC
\citep{rd_92}.
The derived abundances (with 90\% confidence interval in brackets)
from the analysis of the {\it Chandra} grating spectra \citep{zhek_09} 
are given in the third column.
\end{table}

To fit the global RSS model to the data we adopt the following
procedure.  We choose one of the density profiles in the HII
region that provides a good fit to the expansion velocity curve 
(Fig.~\ref{fig:cheva}). For the results presented here we used the
case with $t_{max} = 13,000$ DAE and $\tilde{\rho_0} = 0.0289$ 
($\tilde{\rho_0} = \rho_0/\rho_{Ring}$). We held these parameters
fixed in the fits. Consequently, we know 
the evolution of the emission measure (EM~$\propto n^2 r^3$) and 
ionization time ($n_e t$) of the blast wave, but in relative 
units. We derive and their absolute calibration from the fits. From 
the analytical solution we also know the velocity of the blast wave 
which in turn determines the mean plasma temperature behind this 
shock. For each observation,
we fitted for the average $\bar{\beta}$ which is a free 
parameter. We derive the mean plasma temperature in the 
transmitted shock self-consistently  as described in 
Appendix B in \citet{zhek_09} and we only fitted for the average 
electron temperature in this shock. For each observed spectrum we
fitted both the emission measure and 
ionization age of the transmitted  shock. 
The RSS model also estimates all  the 
parameters (and the spectrum) of the reflected shock in a 
self-consistent way.

\begin{figure*}
 \centering\includegraphics[width=3in,height=2.0in]{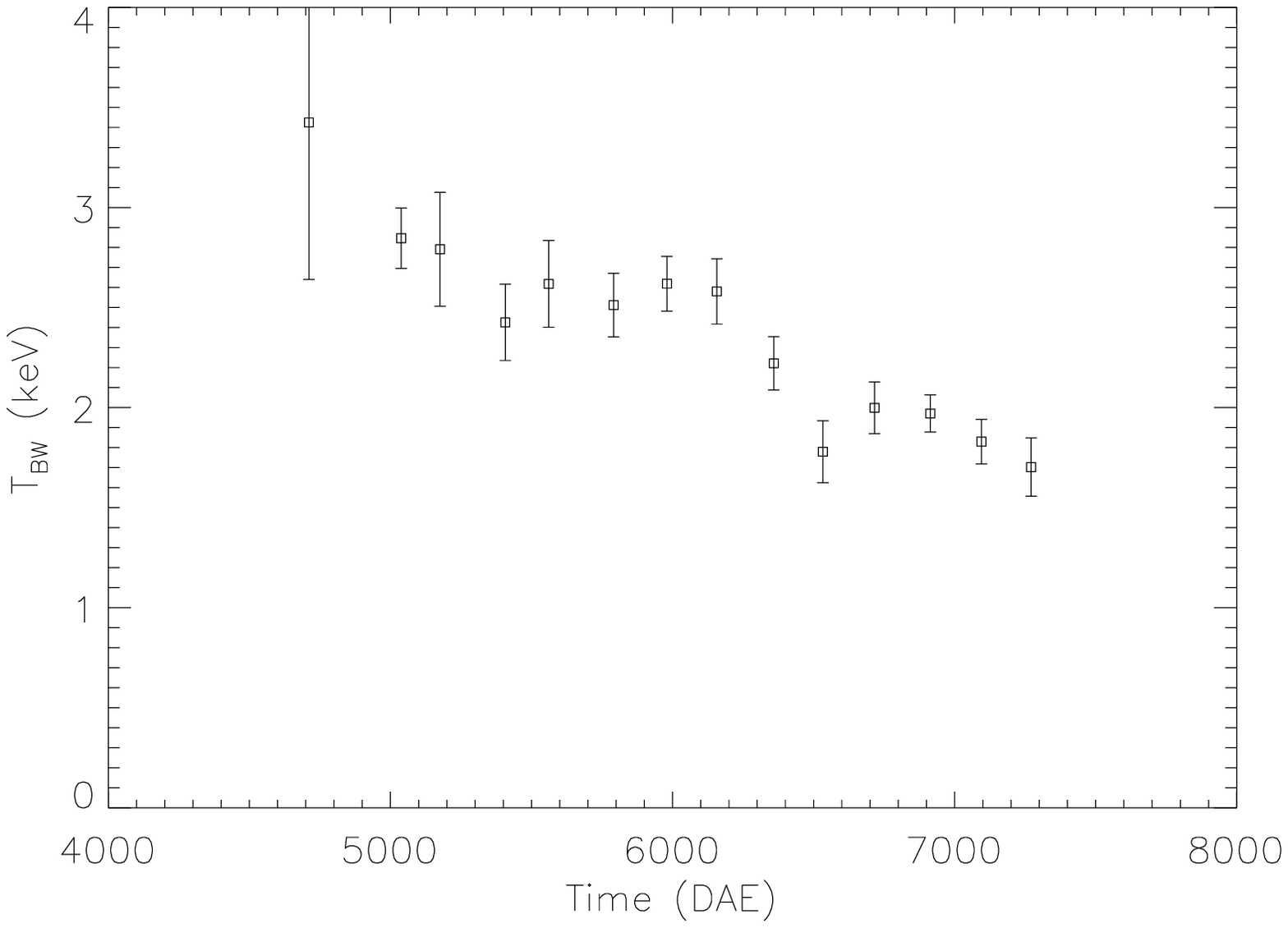}
 \centering\includegraphics[width=3in,height=2.0in]{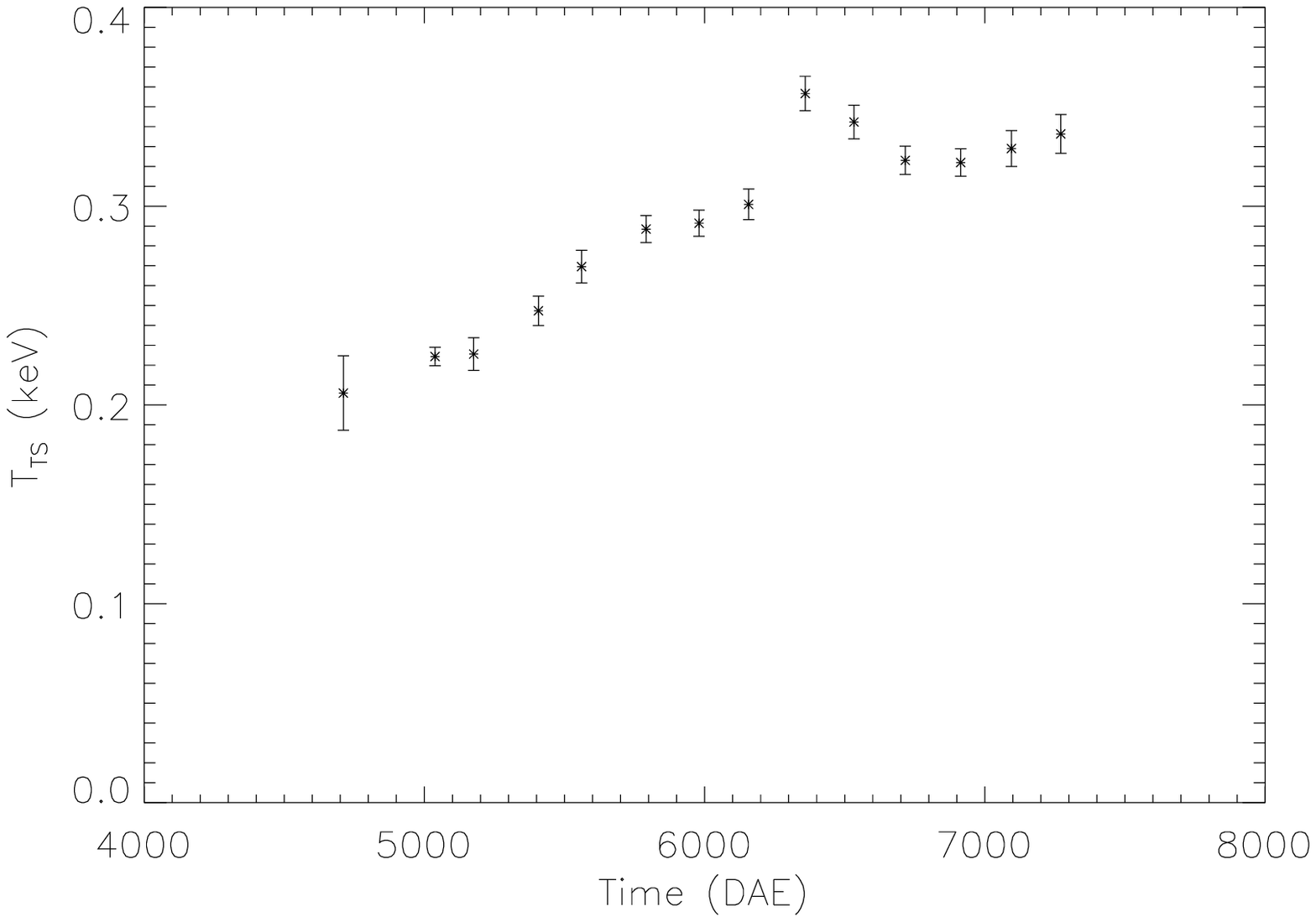}
 \centering\includegraphics[width=3in,height=2.0in]{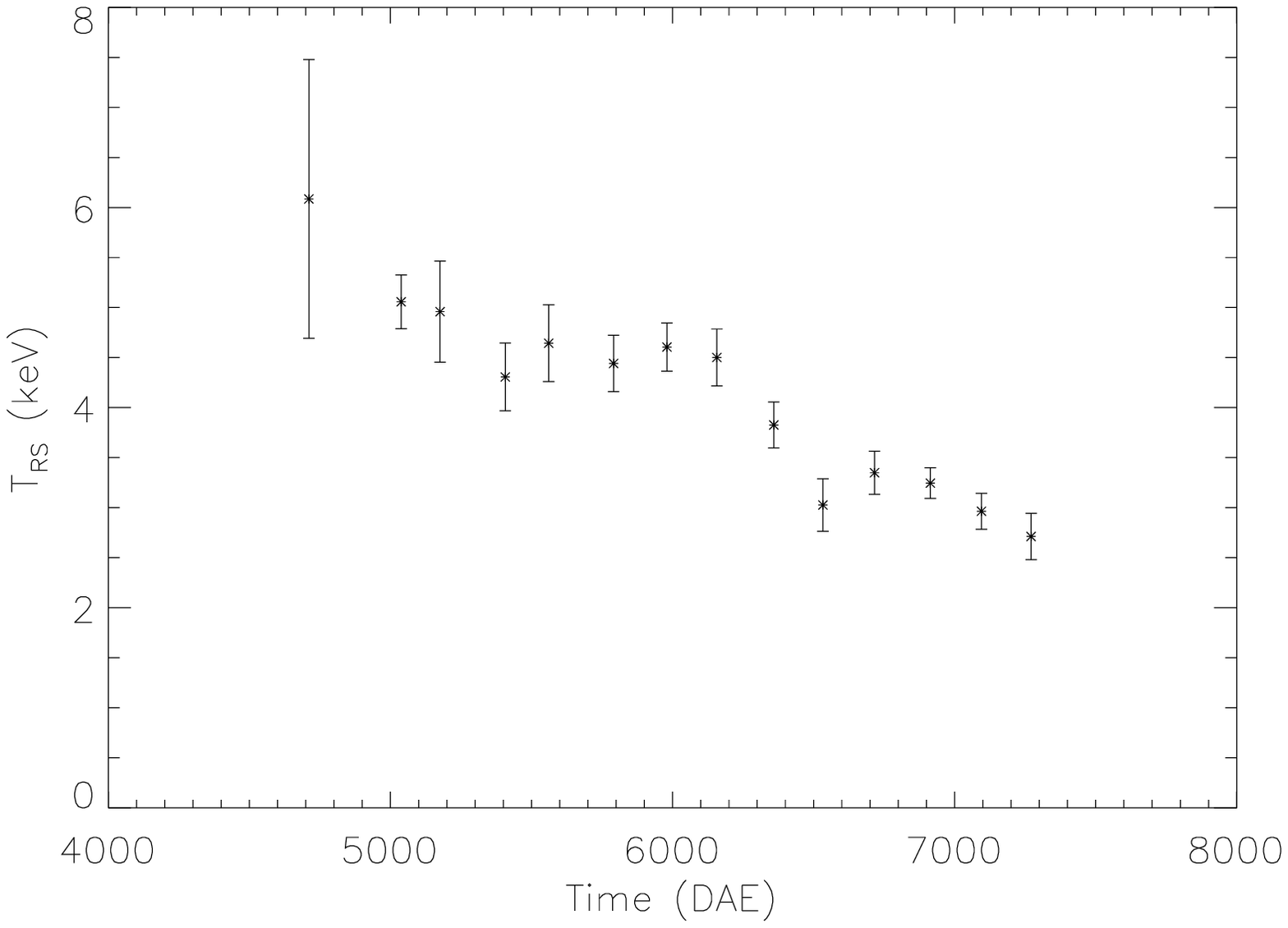}
 \centering\includegraphics[width=3in,height=2.0in]{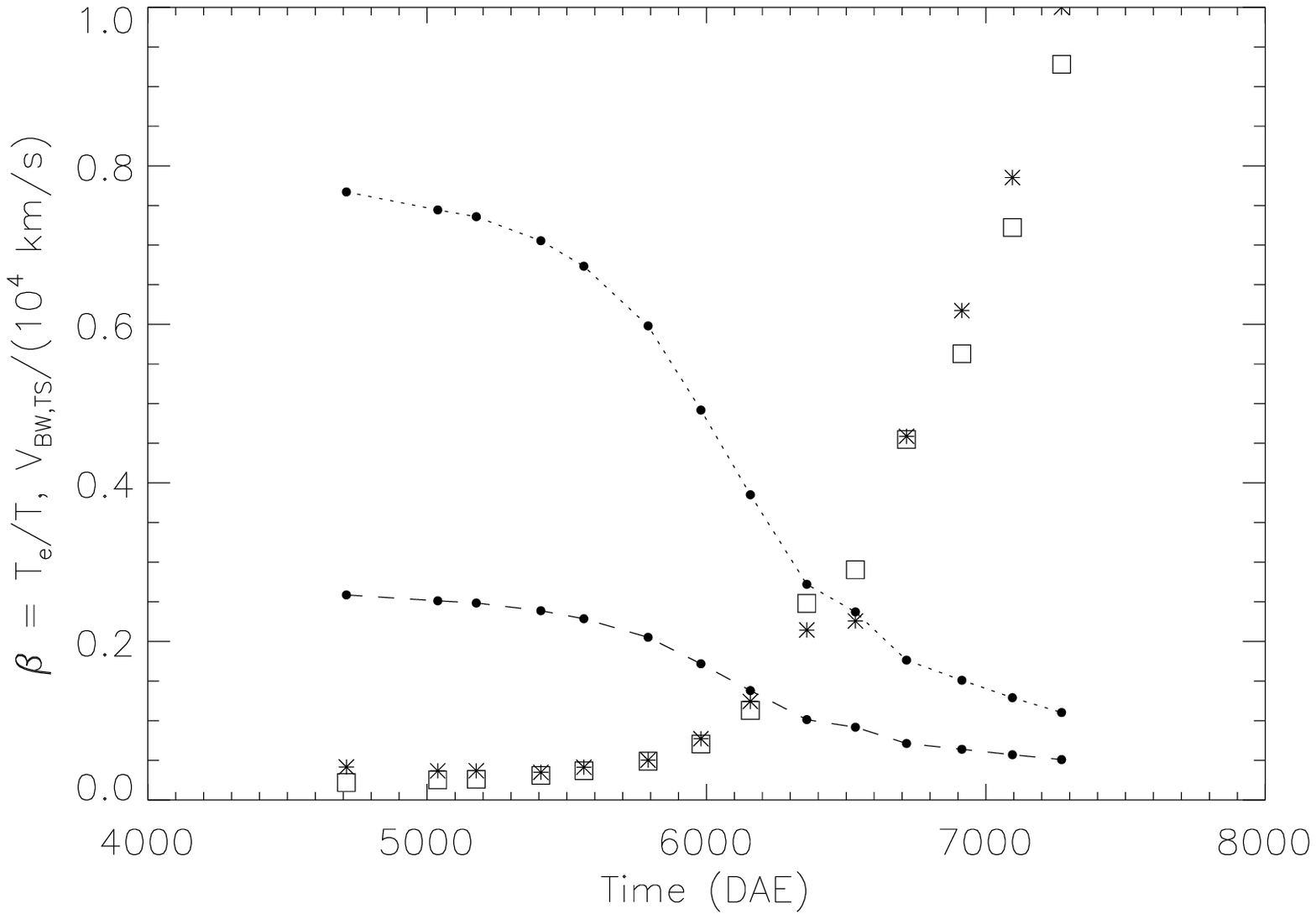}
\caption{
Evolution of various shock parameters as derived from the global
RSS model fit.
Shown are:
the average electron temperatures for the blast wave, transmitted
and reflected shocks (see text for details). The lower right panel
represents the ratio of the average electron to the mean postshock
temperature for the blast wave (squares) and the transmitted shock
(asterisks). For comparison the blast wave velocity is denoted with
dots and a dotted line, while the velocity of the transmitted shock is
depicted with dots and a dashed line.
The error bars in the temperature plots are $1\sigma$ errors from
the fit.
BW, TS and RS stand respectively for the blast wave, transmitted  and
reflected shock.
}
\label{fig:temp_beta}
\end{figure*}

\begin{figure}
 \centering\includegraphics[width=3in,height=2.0in]{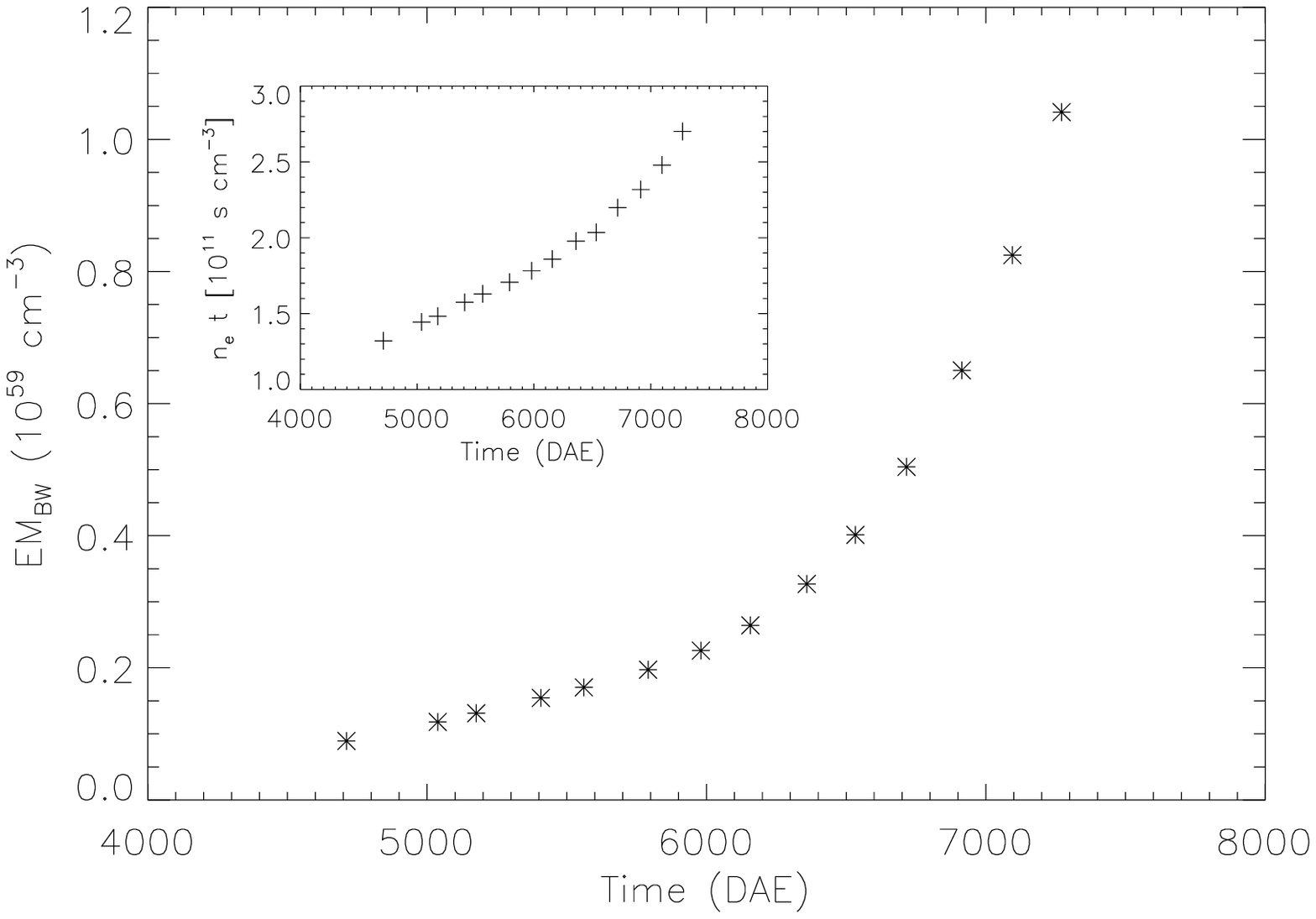}
 \centering\includegraphics[width=3in,height=2.0in]{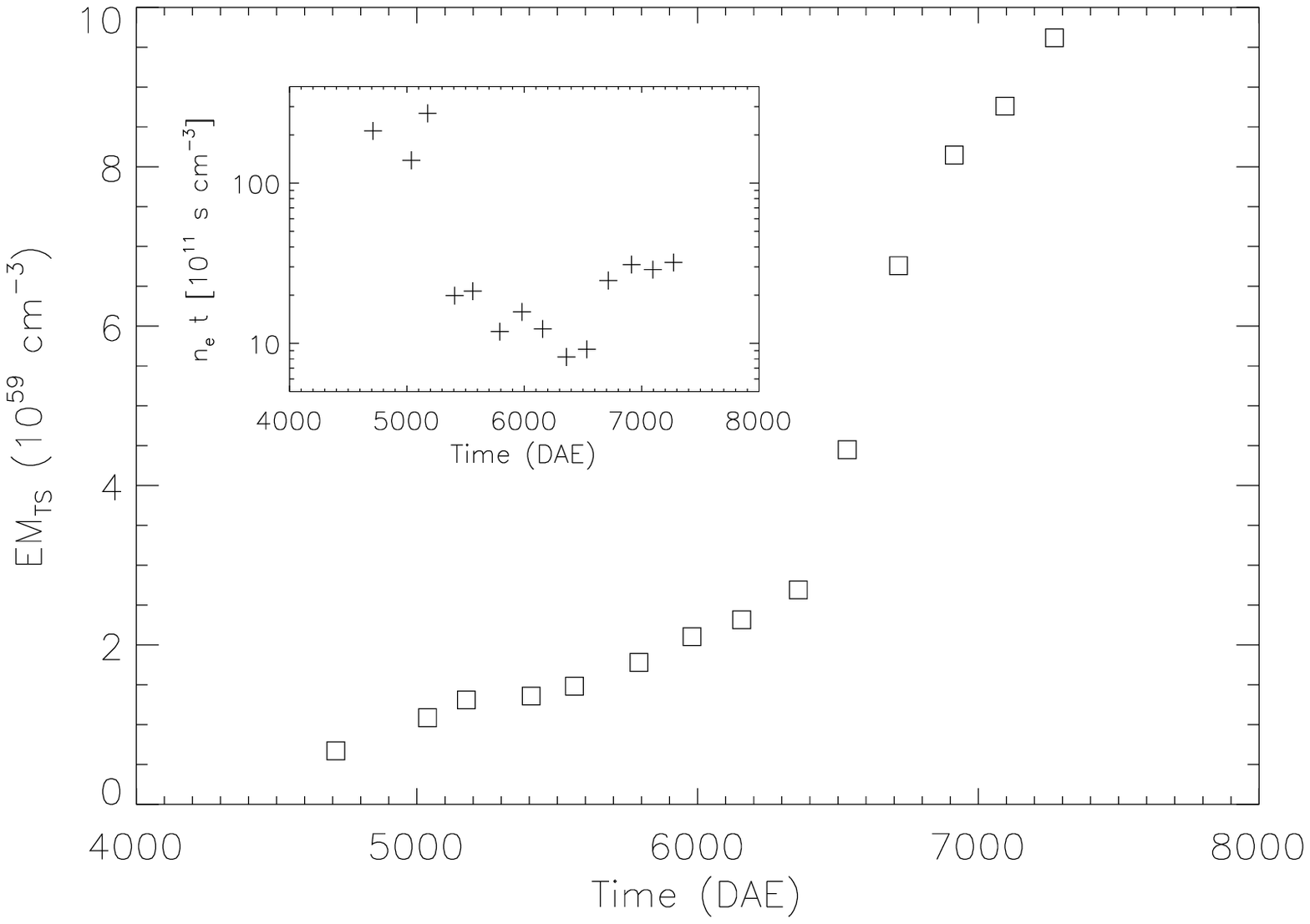}
 \centering\includegraphics[width=3in,height=2.0in]{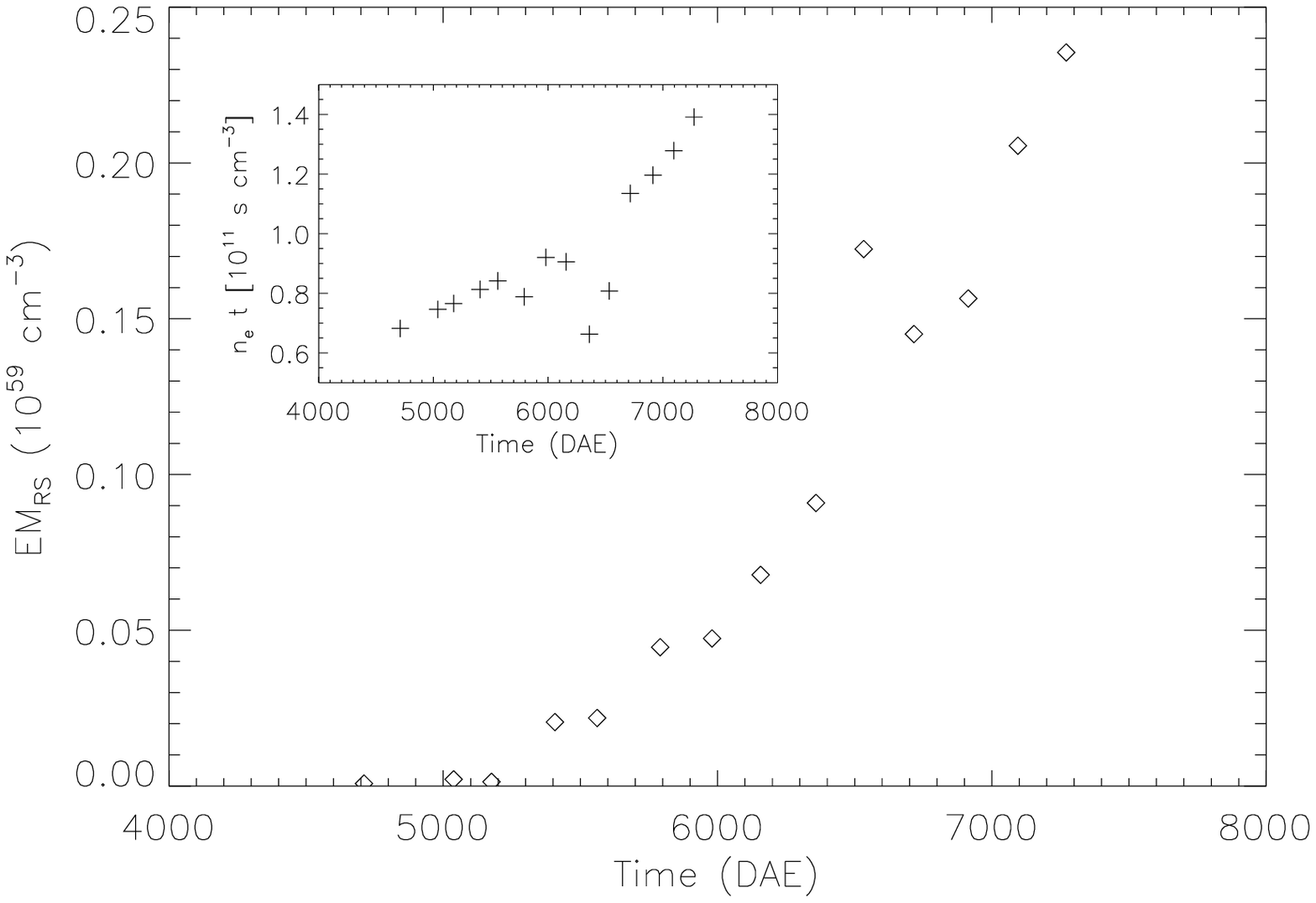}
\caption{
Emission measure of the shock components as derived from the global
RSS model fit.
BW, TS and RS stand respectively for the blast wave, transmitted  and
reflected shock.
The corresponding ionization ages are shown in the inset figure of
each panel.
}
\label{fig:emm}
\end{figure}

The global RSS model was used to fit 14 CCD spectra simultaneously.
Figure~\ref{fig:spectra} shows that
the RSS model is very successful in representing the
X-ray emission from \snrE.  Other fit results are given in 
Table~\ref{tab:fit},  Figs.~\ref{fig:temp_beta} and \ref{fig:emm}.
One can see that some shock parameters follow the same pattern as
derived in previous analyses (e.g. \citealt{sang_06}) but their values
are derived now in the framework of a (highly simplified)
{\it physical} model. For example, the average electron temperature 
in the blast wave decreases gradually, while the mean electron 
temperature behind the transmitted shock increases slightly. 
These results are consequences of the increasing preshock
density and the corresponding decreasing shock velocity, which helps 
the electron-ion temperature equilibration. As 
Figure~\ref{fig:emm} shows, the blast-wave emission measure increases 
smoothly at an accelerating rate. In contrast, the emission
measure of the transmitted and reflected shocks increases suddenly 
for epochs after $\sim 6200~(6000 - 6400)$ DAE in coincidence with
the steep upturn in the X-ray LC  \citep{sang_05} and the
decrease in the expansion velocity curve of \snr \citep{judy_09}.
This behavior is probably a consequence of the fact 
that the number of dense clumps overtaken by the blast wave
started to increase at those times. Future observations will tell
us about their distribution deeper in the
equatorial ring.
Also, Figure~\ref{fig:emm} shows the ionization ages (n$_e$t) 
for each RSS component. We note that the high values for the
transmitted shock in the early observations likely result from
numerical uncertainties due to the small contribution of this
component to the total X-ray emission.
Thus, we consider these (n$_e$t) values less reliable.

\begin{figure}
 \centering\includegraphics[width=3.0in,height=2.0in]{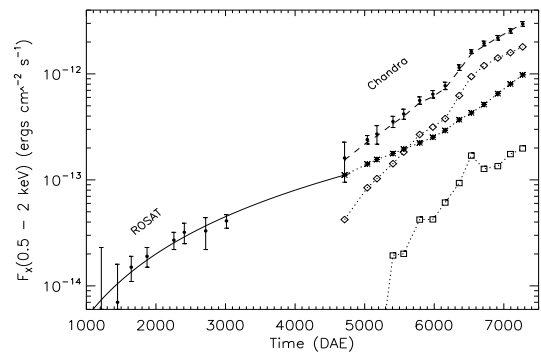}
\caption{
The soft X-ray LC of \snr from {\it ROSAT} and {\it Chandra}.
The data with error bars are from Park et al. (2007).
Components of the RSS model are marked with:
asterisks (the blast wave); diamonds (the transmitted shock);
squares (the reflected shock).
The solid line denotes the extrapolation from the first {\it Chandra}
observation back to the {\it ROSAT} times under the assumption
that all the X-ray emission is due to the blast wave alone.
}
\label{fig:flux1}
\end{figure}

\begin{figure}
 \centering\includegraphics[width=3.0in,height=2.0in]{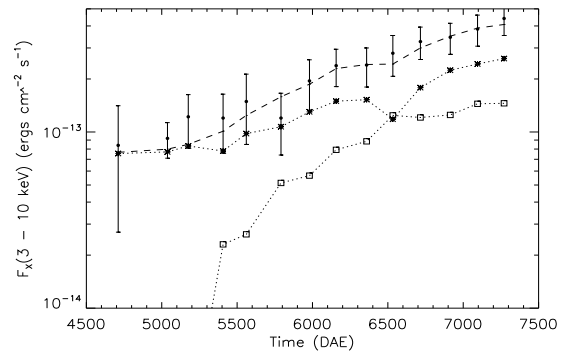}
\caption{
The hard X-ray LC of \snr from {\it Chandra}.
The data with error bars are from Park et al. (2007).
Components of the RSS model are marked with:
asterisks (the blast wave) and
squares (the reflected shock).
The contribution of the transmitted shock is very low and thus it is
out of the shown range on the y-axis.
}
\label{fig:flux2}
\end{figure}

Note that on average the derived abundances have smaller values than
the ones derived from analysis of the grating data (\citealt{dew_08};
\citealt{zhek_09}). This result is likely an artifact of the
diminished resolution in the CCD spectra.
Figures~\ref{fig:flux1} and ~\ref{fig:flux2} show how the RSS model
matches the X-ray light curve in the `soft' (0.5-2 keV) and `hard' 
(3-10 keV) energy bands.
All the fluxes are from \citet{sang_06} and \citet{sang_07}.
We see that the emission from the blast wave dominated the soft X-ray
light curve until the turning point mentioned above, 
$\sim 6200~(6000 - 6400)$ DAE, after which time the emission from the 
transmitted shock became dominated.
On the other hand, the blast-wave emission continued to dominate
the hard X-ray light curve after the turning point, augmented by an
increasing contribution from reflected shocks.

Interestingly, the (0.5-2 keV) model flux, extrapolated back to 
{\it ROSAT} times, matches those \snr fluxes as well. This 
extrapolation assumes that at that time the entire X-ray emission 
was formed in the blast wave. Therefore, the blast wave flux from 
the first {\it Chandra} observation was scaled back with the emission 
measure of the blast wave (EM~$\propto n^2 r^3$) which is known from 
the analytical solution. This result gives us 
more confidence in the global RSS model.

Figure~\ref{fig:temp_beta} shows that
the average $\beta$ ($\beta = \bar{T_e}/T$) for the 
blast wave gradually increases as the shock velocity decreases. 
This behavior is similar to the one for the ratio of the electron
and ion temperature set at the shock front being reversely 
proportional to the Alfv\'{e}n Mach number of the shock 
(e.g. \citealt{ghava_01}). The velocity 
evolution of the average $\beta$ can likely be attributed to this 
fact as well as to the increasing ionization age of the blast wave. 
However, one should keep in mind that the actual physical 
picture is not that simple since any new gas parcel that is overtaken 
by the blast wave does {\it not} follow exactly the evolution of the 
previous gas parcels and has its own evolution instead. 

Note also that the average $\beta$ of the
transmitted shock follows exactly the same pattern as that for the
blast wave. We regard this result  with caution:
it might just be a coincidence that the values of the two 
$\beta$'s are about the same
since 
although relatively high 
the velocity of the transmitted
shock is considerably lower than that of the blast wave.

Summarizing the basic results, we note that the global RSS model 
provides a simple unified physical picture that can match the
observed expansion of the X-ray image as well as the X-ray light curve
and spectral evolution of \snrE.
We are encouraged to believe that by using the RSS model to
analyze
the data from  the ongoing monitoring of this object with {\it Chandra}
we may deduce further details, not only about CSM, but also
about the kinematics of the X-ray emitting plasma.
Moreover, we are hopeful that the information derived from the RSS
model analysis of the X-rays from \snr may be used in conjunction 
with observations at other wavelength bands (e.g., radio, infrared) 
to further constrain the model.

\section{Discussion}
\label{sec:discussion}

\subsection{X-rays vs. Radio Emission}
\label{subsec:xrays-radio}
X-rays from \snr are of thermal origin and they provide us with direct
information about the strong shocks in this object. On the other hand,
the radio emission from \snr is non-thermal synchrotron
radiation (e.g. \citealt{stav_92}), which requires two 
ingredients: relativistic electrons and a magnetic field.
The relativistic electrons 
are assumed to be born in strong shocks through the mechanism 
of diffusive shock acceleration (DSA). Thus, the non-thermal (NT) 
radio emission can 
provide us with additional details about the physical picture in \snrE. 

An interesting question is whether the X-rays and NT radio
emission originate from the same shocks. 
In what follows, we assume that this is indeed the case, specifically, 
that the radio emission comes from the blast wave in \snrE. 
To explore this possibility, we  adopt an approach similar to the one
we took in the analysis of the CCD X-ray spectra above. Namely, we 
adopt a simple model for the NT radio emission and derive some 
parameters of the relativistic electrons involved in this process.
Because it is not possible to measure the magnetic field directly,
we assume that it is proportional to the density of the shocked 
plasma (frozen-in conditions).

\begin{figure*}
 \centering\includegraphics[width=6.in,height=4.in]{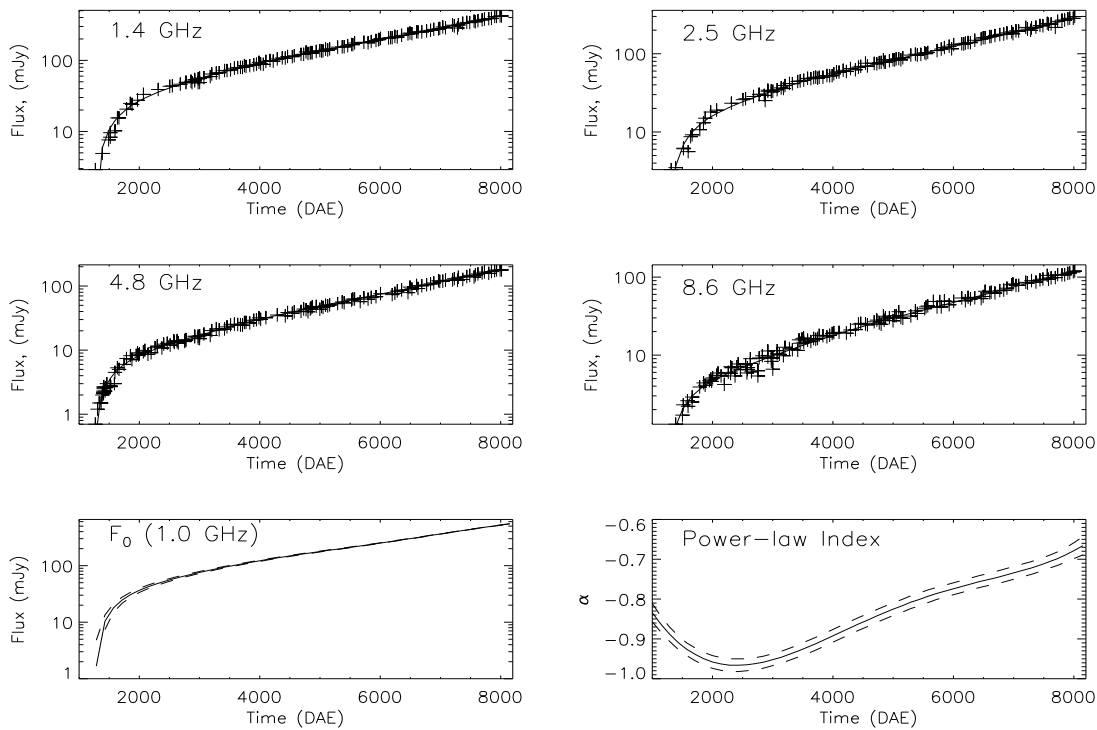}
\caption{
The radio LCs (pluses) of \snr (from \citealt{zana_10}) 
and their model fits with eq.(\ref{eq:fit}) (solid line).
The model parameters, flux normalization at 1 GHz ($F_0$) and
spectral index ($\alpha$), are shown in the bottom two panels.
The dashed lines represent the corresponding $1\sigma$ errors 
from the fit. 
The $rms$ from the entire fit is 3.9 mJy while its value for
the individual frequencies is 4.8 mJy (1.4 GHz),
5.2 mJy (2.4 GHz), 2.8 mJy (4.8 GHz) and 2.4 mJy (8.6 GHz).
}
\label{fig:radio_lc}
\end{figure*}

The light curves (LC) of \snr at four radio frequencies were taken 
from \citet{zana_10}.
Under the assumption that this is synchrotron emission from
relativistic electrons having a power-law distribution in energy, the
LCs were fitted simultaneously with a power-law function whose
normalization and power-law index are function of time:

\begin{equation} \label{eq:fit}
F_{\nu}(t) = F_0(t)\hspace{1mm} \nu_{\mbox{GHz}}^{\alpha(t)}
\end{equation}
where $\nu_{\mbox{GHz}}$ is the radio frequency in GHz, $F_0(t)$ is 
the flux at 1 GHz in mJy and $\alpha(t)$ is the power-law index. 
The parameters
$F_0(t)$ and $\alpha(t)$ were characterized as a sum of Chebyshev
polynomials of up to the 6-th power: 
$F_0(t) = \sum a_j~Cheb_j$, $\alpha(t) = \sum b_j~Cheb_j$.
We note that such a characterization allows deriving the power-law
index even if the data at different frequencies are not taken exactly
at the same time.

The fits to the radio LCs and the derived flux normalization ($F_0$)
and power-law index ($\alpha$) are shown in 
Fig.~\ref{fig:radio_lc}. Note that the variations of the power-law 
index are similar to those derived in the original analysis of
these data (\citealt{man_02}; \citealt{stav_07}; \citealt{zana_10}). 
Next we interpret these results by considering the basic model
\citep{bell_78} for diffusive shock acceleration of relativistic
electrons and the theory of synchrotron emission \citep{rl_79}.

In the framework of the DSA model, the resultant relativistic particles 
have a power-law distribution in energy and the corresponding power-law
index is (eq. [12] in \citealt{bell_78}):

\begin{equation}
p = \frac{(2 + d_{ju}) + d_{ju} (2 V_w/V_{sh} - 1/M_A)}
{(d_{ju}- 1) - d_{ju} (V_w/V_{sh} + 1/M_A)}
\end{equation}
where $d_{ju}$ is the density jump at the shock front, $V_{sh}$ is the
shock velocity, $V_w$ is the velocity of the turbulent waves
downstream and $M_A$ is the Alfv\'{e}n Mach number of the shock. We 
note that for strong adiabatic shocks (as expected in \snrE) the 
density jump is $d_{ju} = 4$ for gas (plasma) with adiabatic index 
$\gamma = 5/3$. Thus, three limiting cases are possible:
(i) shocks with large  Alfv\'{e}n Mach number and turbulent 
waves which have quickly isotropised ($V_w \ll V_{sh}, M_A \gg 1$);
(ii) strong shocks with similar turbulent waves as above but with
Alfv\'{e}n Mach number not much larger than unity ($d_{ju} = 4,
V_w \ll V_{sh}, M_A > 1$);
(iii) strong shocks with large Alfv\'{e}n Mach number and turbulent
waves that have not isotropised yet ($d_{ju} = 4, V_w/V_{sh} \neq 0;
M_A \gg 1$).

\begin{figure*}
 \centering\includegraphics[width=6.in,height=4.in]{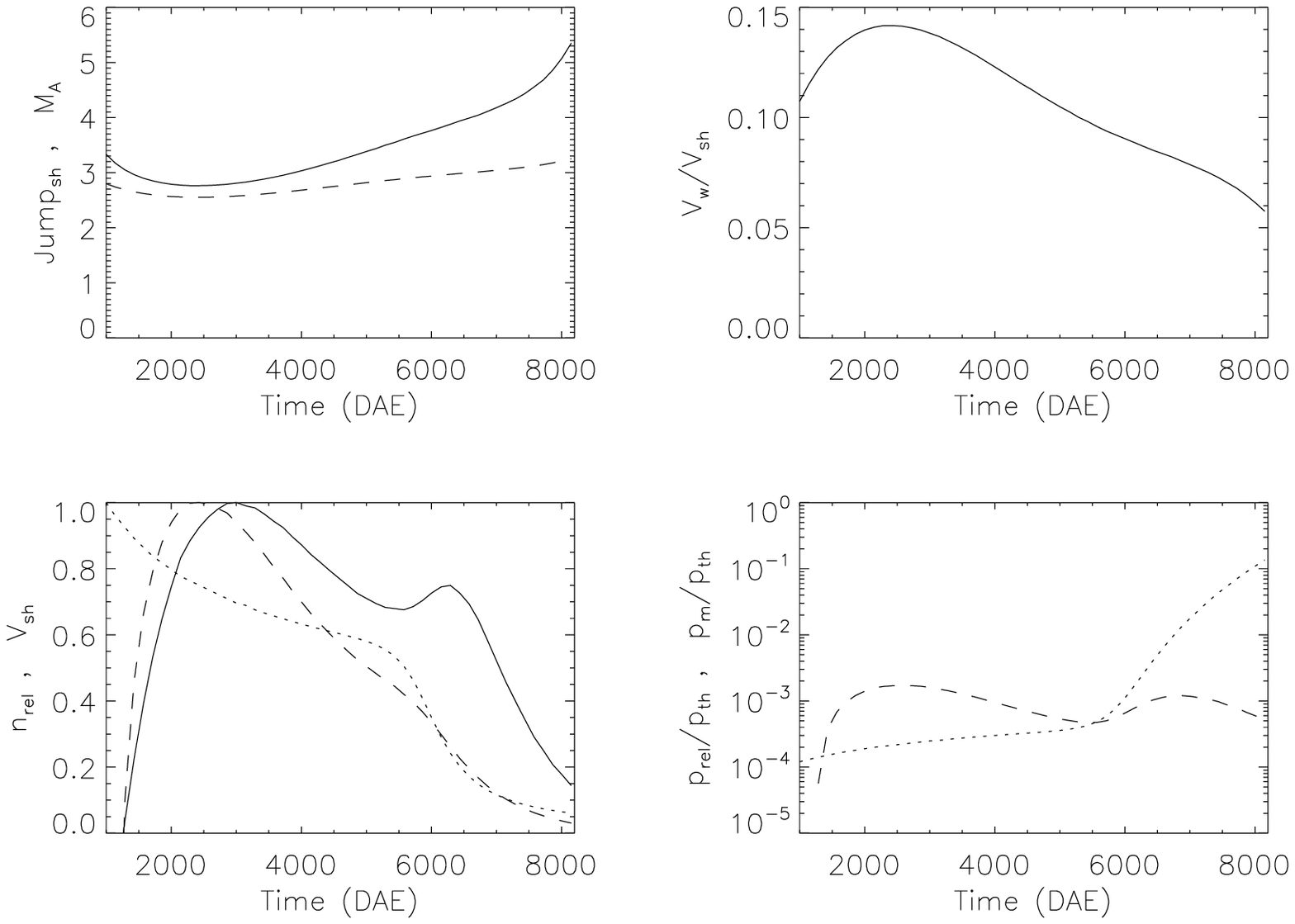}
\caption{
The radio LC fit parameters of \snrE.
The {\it Upper  Left} panel:
the Alfv\'{e}n Mach number (solid line) and the density jump (dashed
line).
The {\it Upper  Right} panel: the turbulent velocity.
The {\it Lower  Left} panel:
the time-averaged production rate of relativistic electrons
(dashed line), the number density of the injected relativistic
electrons (solid line) and the blast wave velocity (dotted line) as
all
these are normalized to their maximum value for clarity.
The {\it Lower  Right} panel: the ratios of the pressure of
relativistic electrons (dashed line) and magnetic pressure (dotted
line) to the thermal pressure behind the blast wave.
See text for details.
}
\label{fig:radio_stuff}
\end{figure*}

The corresponding results for these three limiting cases are shown in
the two upper panels of Figure~\ref{fig:radio_stuff}. We made use of
the relation between the spectral index of the resultant
synchrotron emission from relativistic electrons with a power-law
distribution in energy, $\alpha = -(p-1)/2$ \citep{rl_79}.
In case (i), we see that the density jump at the shock(s) in \snr is
smaller than its canonical value ($d_{ju} = 4$) for strong shocks.
This may indicate that efficient particle acceleration influenced the
hydrodynamic structure of the shock(s). If so, the derived values for 
the density
jump correspond to the subshock while the total compression
(for the downstream versus upstream gas) is greater than 4 
(e.g., for the case of \snr see  \citealt{ber_ksen_06}; 
\citealt{duf_95}).
In case (ii), we see that the Alfv\'{e}n Mach number has a tendency to
increase with time. We note that the opposite is expected if the blast
wave penetrates further on in the HII region near the inner ring and
the magnetic field is subject to the frozen-in conditions (being
proportional to the density).
In case (iii), we see that after an initial increase the turbulent waves 
velocity is decreasing which may indicate that they evolve towards
equilibrium --- that is they tend to isotropize with time.

On the other hand, the fits to the radio LCs of \snr tell us 
the time evolution of the normalization for its flux ($F_0$) which
in turn can be used to deduce some valuable information about the
relativistic electrons, magnetic field etc. From equation (6.36) in 
\citet{rl_79} and for a distance of 50 kpc to \snr we can
write:

\begin{equation}\label{eq:norm}
F_0 = 4.272\times10^{11} \frac{n_{tot}^{(r)} \tilde{V} B^{1-\alpha} 
\Gamma_1\Gamma_2 C_0^{\alpha} (-2\alpha)}{2(1-\alpha)}
\end{equation}
where $F_0$ is in mJy;
$n_{tot}^{(r)}$ is the total number density of the relativistic
electrons;
$\tilde{V} = V(t)/V_R = (r/R_{Ring})^3 \tilde{\Delta}$ is the 
volume occupied by the shocked gas (and relativistic electrons), that 
is between the blast wave and the contact discontinuity, in relative units
and $\tilde{\Delta} = \Delta r / r$ is the relative thickness of that
volume; 
$B$ is the magnetic field strength in Gauss 
($B = B_0\frac{n}{n_0}$ - frozen-in conditions, the subscript $0$
denotes parameters at the inner radius of the HII region); 
$C_0 = (2\pi m_e c/3e)\times 10^9 =  119.119$ is a constant ($m_e,
e$ are the electron mass and charge,  $c$ is the speed of light, while 
the factor $10^9$ comes from using the radio frequencies in GHz);
$\Gamma_1 = \Gamma(\frac{p}{4}+\frac{19}{12})$ and 
$\Gamma_2 = \Gamma(\frac{p}{4}-\frac{1}{12})$ as $\Gamma$ is the gamma
function.

Thus, from eq.~(\ref{eq:norm}) and the fit results for the radio LCs,
we can derive the time evolution of the mean number density of
relativistic electrons. We recall that the radius of the blast wave,
and the preshock density profile are those used in the fits to the X-ray
CCD spectra of \snrE. The relativistic number density can then be used
to estimate the pressure of relativistic electrons in the 
radio-emitting region.
Figure~\ref{fig:radio_stuff} presents results for a nominal
set of values for the preshock nucleon density and magnetic field
at the inner radius of the HII region ($n_0 = 100$~cm$^{-3}$ , 
$B_0 = 1$~mG) and thickness (e.g., $\tilde{\Delta} \approx 0.1$).
We recall that the gas pressure behind the shock is $p_g = 3/4 \rho_0
V_{sh}^2$ ($\rho_0 = 1.6 m_p n_0$ for the \snr abundances, 
$m_p$ is the proton mass).
We see that neither the pressure of relativistic electrons nor the 
magnetic field pressure is a significant fraction of the thermal gas 
pressure. 
This indicates that the model of standard strong shocks can explain 
both the observed X-ray and radio emission from \snrE. Also, the small 
magnetic pressure is a sign of large Alfv\'{e}n Mach numbers, thus, 
case (iii) mentioned above seems a realistic option for the
corresponding physical picture,
and future observations will show us if this picture may evolve to
being intermediate between cases (ii) and (iii).

From eq.~(\ref{eq:norm}) and the fit results for the radio LCs, we can 
also estimate the time evolution of the total number of
relativistic electrons in the shock region and some related quantities.
Namely, the total number of particles is $n_{tot}^{(r)} \tilde{V} =
N(t) = N(t_0) + \bar{\dot{N}}(t) (t - t_0)$, where $\bar{\dot{N}}(t)$ 
is the time-averaged production rate of relativistic electrons.
Since the relativistic particles are accelerated at the shock front
and then injected into the postshock region, we could also write:
$N(t) = N(t_0) + n_{in}^{(r)} u_2 S \Delta t$, where $n_{in}^{(r)}$ is
the number density of the injected relativistic electrons, $S$ is
the shock surface ($S \propto [r/R_{Ring}]^2$) and according to the 
DSA model: $u_2 = V_{sh}/4 + V_w$ for strong shocks \citep{bell_78}.
Figure~\ref{fig:radio_stuff} presents the time evolution of $N(t)$ and
$n_{in}^{(r)}$ based on the fits to the radio LCs and for case (iii).
We see that after an initial steep increase of the newly born
relativistic electrons, their average production rate follows a
gradual decline. 
Note also that the number density
of the newly injected relativistic electrons, normalized to the
local preshock density, follows a similar pattern.
We tentatively attribute this decline to decreasing efficiency of
particle acceleration in the decelerating shocks.
We believe that this behavior may be an important clue
to the details of the particle acceleration mechanism
that operates in \snrE. Such details might be helpful to build a
physically more sophisticated  model of this system.

\subsection{Overall Consistency of the Global RSS Picture}
\label{subsec:overall}
As a further constraint on this picture and as a check on its internal
consistency, we can use the results from the fits to the CCD X-ray 
spectra of \snrE. Namely, the preshock nucleon number density can be
deduced from the emission measure of the blast wave and a density jump
for strong adiabatic shocks ($d_{ju} = 4$). For distance of 50 kpc to
\snr the X-ray fits yield a nucleon number density at the inner radius of
the HII region: $n_{0, X} = (49.2\pm1)/\sqrt{\sin\theta}$, cm$^{-3}$,
where $\theta$ is the opening half-angle of the inner ring
($\theta = \pi/2$ gives a complete spherical shell). 
Recall that we have a clear evidence based on the spatial-spectral
effects in grating spectra that the X-ray emission in \snr is confined
in the equatorial plane  although the opening angle of the X-ray
emission region is not yet well constrained (\citealt{zhek_05}, 2006; 
\citealt{dew_08}; \citealt{zhek_09}). On the other hand, the
triple-ring system  observed in the optical \citep{burr_95}
is solid evidence for a highly asymmetric distribution of the 
circumstellar matter around the exploded star in the center of \snrE.
Although the
origin of this asymmetry is not yet well understood, the scenario of
colliding stellar winds that considers a spherically symmetric BSG
wind impinging onto an asymmetric RSG is probably the most realistic 
model so far. \citet{bl_lu_93} investigated this scenario
in detail and their `best' model, which correctly predicts various
parameters of the circumstellar environment prior to the explosion of
SN~1987A, requires that 50\% of the RSG wind mass is collimated
within $\sim 10^{\circ}$ from the equatorial plane. We thus see that
the nucleon density of the HII region that is deduced from the
analysis of the CCD X-ray spectra is $n_{0, X} = 118\pm2.5$, cm$^{-3}$.
This value is consistent with the one assumed for the estimates
presented in Fig.~\ref{fig:radio_stuff}.

From these considerations we conclude that the case of strong adiabatic
shock(s) is suitable for explaining both the X-ray and radio emission
from \snr and that further analysis of the newly obtained data along 
the
same lines might be indeed helpful in building a realistic picture of
this object. It is worth noting, though, that this is indeed the case
provided the DSA of heavy particles is not as efficient as
the one for the electrons in \snrE. For example, if the relativistic 
protons gain the same amount of energy from the blast wave as 
supplied into relativistic electrons the picture will remain the same.
But, if the DSA is very efficient and produces relativistic
protons (or nuclear cosmic rays) and electrons with similar densities, 
the total pressure
of all the relativistic particles will be comparable to the thermal
gas pressure and the physical picture will be more similar to case (i)
as described above. We recall that in such a case the blast wave in 
\snr will be affected by the feedback from the relativistic particles
which will result in changed physical parameters in the postshock
region. A self-consistent treatment of these effects is needed which is
beyond the scope of this study. We only note that the kinematics of
the postshock (X-ray emitting) plasma will be influenced most, thus,
a continuing monitoring of the X-ray emission from \snr with high
spectral resolution (grating spectra) will be crucial in this respect.

Some deceleration in the bulk gas velocities
was found between 2004 and 2007 from the grating spectra with very
good photon statistics. The most important kinematic result from 
these data, however, was the finding that the bulk gas velocities 
inferred from the X-ray 
line profiles are too low to account for the postshock plasma 
temperatures derived from the spectral fits (\citealt{zhek_05}, 2006;
\citealt{dew_08}; \citealt{zhek_09}). 
On the other hand, if
efficient DSA operates in \snrE, the total gas compression downstream
from the shock front (i.e., in the X-ray emitting region) will be 
greater than its canonical value ($d_{ju,tot} > 4$). Such a result
would imply that the bulk gas velocities in the space will be greater:
$V_{bulk} = \frac{d_{ju,tot}-1}{d_{ju,tot}}V_{sh}$ (e.g., see also 
\citealt{cheva_83}).  If so, the X-ray spectral lines would be
broader, in conflict to observations. These considerations support 
our assumption that the DSA in \snr is not very efficient.

This assumption gets additional support from the fact that
the X-ray spectrum of \snr  is dominated by emission
from thermal plasma. 
Moreover, if, using the results from the fits to radio LCs 
($F_0, \alpha$),
we extrapolate the NT radio spectrum to X-rays (equivalent to
assuming efficient particle acceleration), the observed flux 
in the (3 - 10 keV) energy band is smaller than the one expected 
from relativistic electrons for the \snr age larger than 5,500 DAE. 
This result confirms that the DSA is not very efficient in producing
relativistic particles with very high energies.

In the analysis presented here, we have argued that the physical
picture in which the X-ray and NT radio emission from \snr originate 
in one and the same shock structure is realistic and capable of
describing the observations. But, there is one important detail 
that emerges 
from the analyses of the X-ray and radio images of this object 
and needs more attention. If the picture 
adopted here is correct, the LCs in X-rays and radio may 
differ but the angular size of \snr in these 
spectral domains must be about the same. However, 
the image analyses so far indicate that
the X-ray size of \snr is consistently smaller than its radio size.
Because the object size is not measured directly but only
after applying image-analysis techniques, such a difference might
be attributed to the different methods used in analyzing the 
X-ray and radio images \citep{gaensler_07}. On the other hand, 
a recent and more elaborate analysis of the radio images of \snr 
\citet{ng_08} find that for times beyond 6,000 days after the 
explosion (6,000 - 7,600 DAE) the radius of the radio remnant is 
greater than 0.83 arcsec, {\it i.e.}, greater than the radius of the
inner ring (see their Table 3). For the same period of time, the 
X-ray remnant still lies within the equatorial ring \citep{judy_09}.

Such an evolution of the radio size of \snr could be indeed important 
since it may indicate that the blast wave has propagated beyond the 
equatorial ring or the radio emission has some contribution from 
regions above and below it. 
The former was concluded by \citet{ng_09} from their analysis of a 
single {\it Chandra HRC} (High Resolution Camera) image taken at
7736 DAE. They also derived a
larger X-ray radius of \snr compared to the results from
\citet{judy_09}. But
if this were the case, the radio (and or X-ray) emission should
have started to decline similarly to the radio (X-ray) evolution soon 
after the explosion when the supernova ejecta was interacting with the
BSG stellar wind. This is to be expected in the framework of the
colliding stellar wind scenario for the origin of the ring system in
this object because the gas beyond the inner ring is that produced
by the wind of the red supergiant progenitor of \snrE.
Therefore, the emission measure in radio 
and X-rays must decrease with time (preshock number density will 
follow $1/r^2$ dependence). No indications of flux decline are seen 
both in the radio and X-ray LCs (e.g., see \citealt{ng_08};
\citealt{zana_10}; \citealt{judy_09}).
And in order to explore the possibility for contribution 
from shocks above and below the equatorial plane, we may need to 
change our currently used geometrical models. Namely, models that 
assume deviation from a simple geometry of a torus, equatorial belt 
etc. might be a more realistic presentation of the actual interaction 
of the blast wave with the inner ring. We believe that the improving
quality of the X-ray and radio data allow for adopting such models
which in turn may provide us with valuable details about the physics
of the interaction phenomenon.

An interesting aspect of our study is about what we can expect in 
the future, provided the physical picture adopted here is correct. 
As mentioned earlier (see \S~\ref{subsec:anal_sol} and 
Fig.~\ref{fig:cheva}), the expansion velocity curve is successfully 
fitted for density profiles that may have different density contrast 
between the HII region and the maximum density of the smooth component 
at the radius of the inner ring. This results in a considerably 
different time ($t_{max}$) when the blast wave will reach the radius 
of the equatorial ring and the X-ray (as well as the radio) emission 
will be at its maximum. We fitted all 14 CCD spectra simultaneously 
with the global RSS model for the cases with $t_{max}$ between 
10,000 and 15,000 DAE (with an increment of a 1000 DAE). These fits 
had similar quality ($\chi^2 = 1696 - 1734$) as our basic case 
($t_{max} = 13,000$~DAE) discussed here in detail. As anticipated, 
it is not possible to discriminate between these cases just based on 
the available data and only future studies of the expansion velocity 
curve and other X-ray properties will help us in this respect. But 
it is worth mentioning that all these fits converge in the 
following.

Since the absolute value of the emission measure of the blast wave is 
constrained by the observed X-ray flux, in all the cases the nucleon 
number density near the inner radius of the HII region had values that
are about the same: $n_{0,X} \approx 100$~cm$^{-3}$ (as before we 
assumed an opening half-angle of $\sim 10^{\circ}$). From this it
follows (see Fig~\ref{fig:cheva}) that the maximum density of the 
smooth component will be $n_{Ring} = 10^3 - 10^4$~cm$^{-3}$ for the 
cases with $t_{max}$ between 10,000 and 15,000 DAE. On the other hand, 
in all these cases we need to have dense clumps with about the same 
typical number density ($n_{clump} \approx 10^4$~cm$^{-3}$) in order 
to match the observed X-ray spectra. We note that the clump density 
(in fact its contrast to the smooth component) controls the 
properties of the transmitted shocks (low temperature plasma). 
Interestingly, an indication of similar preshock number density 
($\sim 10^4$~cm$^{-3}$) was found also for the first hot spot
discovered in the optical and UV. Using the line ratios of various
forbidden lines, \citet{pun_02} showed that the optical emission 
of Spot 1 comes from regions with densities of 
$n_e \sim 10^6$~cm$^{-3}$ downstream radiative shocks that compressed
the preshock gas by a factor of $\ge 100$.
Thus, 
it seems conclusive that the HII region consists of dense clumps with 
similar properties and the density of its smooth component is not well 
constrained but it will be so depending on the time when the X-ray 
emission reaches its maximum.

It is worth noting that although our global RSS model is successful
in matching both the expansion velocity curve of \snr and the
evolution of its X-ray emission, it
neglects the role of the reverse shock for the latter.
In fact,
the emission from the entire two-slab region (see
\S~\ref{sec:rss_picture}), that is of the shocked material between the
forward and reverse shocks, is represented by the X-ray emission
behind
one fast shock: the evolving blast wave interacting with dense clumps.
The validity of this seems justified given the good fits of our model 
to the X-ray spectra. So it could well be that the reverse shock 
produces unusually cool plasma, as it is a shock going through an
almost neutral medium, in which magnetic turbulence, needed for 
collisionless shock heating, is damped. However, there are enough 
uncertainties in our model, but also concerning the general knowledge 
about collisionless shocks to remain cautious.
We only note that the physics of the reverse shock in \snr is quite 
complicated to be addressed properly in a simplified physical picture 
we have adopted in this work.

As indicated by the infrared observations (e.g. Fig. 18 in
\citealt{bou_06}) , the SN ejecta has become
neutral already at times before
the \snr reappearance in the radio and X-rays ($\sim 1,000$~DAE). 
On the other hand, detection of H$\alpha$ and Ly$\alpha$
emission from the reverse shock (e.g. \citealt{eli_03};
\citealt{heng_06}) is a clear sign for presence of both neutral and
ionized species. Thus, the reverse shock in \snr is not a
standard collisional or collisionless shock in {\it
plasma} but instead its shock front likely consists of a relatively
broad transition zone where the ionization state changes from neutral
(preshock ejecta material) into ionized (compressed ejecta plasma).
As a result, it might well be that the electrons are not preheated 
in the shock front and their temperature is determined by the slow 
Coulomb collisions with protons and ions downstream the reverse shock 
which may not result in strong X-ray emission.

But, there is an interesting transformation of the reverse shock that
we may witness in the near future. As suggested by \citet{nathan_05},
the rapidly increasing X-ray flux from \snr should reach a level
beyond which it will be capable of ionizing the neutral gas before it 
crosses the reverse shock. When this happens, the broad H$\alpha$ and 
Ly$\alpha$ emission will vanish and most importantly the reverse shock 
will transform from its current state into a collisionless shock, 
provided a magnetic field is present in the supernova ejecta.

The reverse shock transformation will have an interesting impact on 
the radio emission from \snr as well. 
\citet{man_05} have suggested that the radio emission originates
from the region behind the reverse shock. Park et al. (2005, 2007)
noticed the close similarity in the shape of the light curves in
the hard X-rays and radio and suggested that they may have a common
origin. But, no signs of non-thermal X-ray emission form \snr have been 
found so far and we argued here that the global RSS picture is
capable of explaining the radio and X-ray properties of \snr by
assuming that the radio emission originates from the region behind 
the blast wave and not behind the reverse shock. 
On the other hand, once the supernova ejecta gas has been ionized
before entering the reverse shock, the shock will become collisionless
which will result in an efficient DSA of relativistic particles. After
that moment, we can expect a more rapid increase of the non-thermal
emission from \snr.

We note that although this picture is only qualitative and thus bears
many uncertainties if it is correct,
two observational evidences should be found in the future that
can serve as a test for its validity.  And it is important to note 
that these events must happen almost {\it simultaneously}: (i) the broad 
H$\alpha$ and Ly$\alpha$ emission from \snr will vanish; 
(ii) there will be a turn-up in 
the non-thermal radio emission from this object.

\subsection{\snr vs. Young Supernova Remnants}
\label{subsec:sn87a-young}
Many young supernova remnants (Cas A, Kepler, Tycho, SN 1006, RCW 86) 
have been observed to contain non-thermal X-ray emission (e.g.
\citealt{all_99}).
These young remnants share other common features:  non-thermal radio 
emission, signs of efficient (nuclear) cosmic ray (CR)  acceleration 
and as a result from that they likely have amplified magnetic fields 
(for a discussion see Volk et al. 2005; Vink 2008)
We note that the key signature of efficient DSA of relativistic 
particles in these objects is the presence of {\it non-thermal} 
X-ray emission. 
As already mentioned (\S~\ref{subsec:overall}), \snr does {\it not} 
have that (or it is very weak) and its X-ray spectrum is successfully 
represented by thermal 
emission from shock heated plasma as analyses of both {\it Chandra} 
(\citealt{eli_02}; Park et al. 2002, 2004, 2006; Zhekov et al. 2006, 
2009)  and {\it XMM-Newton} (\citealt{haberl_06}; \citealt{heng_08}) 
spectra have shown. This is also confirmed by our current analysis
of the evolution of the {\it Chandra} CCD spectra of \snr.

It is then suggestive that the lack of non-thermal X-ray 
emission from \snr could be a signature of inefficient DSA of 
relativistic 
particles (protons and nuclear CR) and it might simply be explained 
by the fact that \snr is very young. Actually, its age is more than one 
order of magnitude smaller than that of the young SNRs 
showing those common features mentioned above. We can speculate that 
some time is probably needed before a SNR reaches a level when the DSA 
becomes very efficient and \snr has not yet entered that stage.
On the other hand, old SNRs are not expected to have strong non-thermal
X-ray emission since their shock velocities are deccelerating (e.g.
\citealt{volk_05}). This can be an alternative explanation for the case 
of \snr that shows clear signs of shock decceleration at times 
$> 5,500$ DAE (\citealt{judy_09}; see also Fig.~\ref{fig:cheva} and the 
lower right panel in Fig.~\ref{fig:temp_beta}).
So,  due to the presence of the inner equatorial ring and its HII
region (a CSM with increasing outward density), the very young
object, \snr, might mimic the evolution of old SNRs.
But in either case, the fact of missing (or very weak) non-thermal 
X-ray emission probably indicates inefficient DSA in \snr.

Thus, the assumption of not very efficient  DSA was adopted in our 
analysis here which also means that the magnetic field strength cannot 
be derived in a self-consistent manner as can be done for young SNRs 
with non-thermal X-ray emission (e.g. Volk et al. 2005). Due to this,
our estimates (see \S~\ref{subsec:xrays-radio} also the lower panels 
in Fig.~\ref{fig:radio_stuff}) use some fiducial value
for the magnetic field strength at the inner radius of the HII 
region and we adopt the frozen-in conditions. The latter means that the
field strength increases deeper in the inner ring.
We note that in the interacting wind scenario, the origin of the
magnetic field is likely related to the global field of the progenitor 
star during the BSG (and RSG) phase of its evolution. The adopted here 
value (B$_0 = 1$~mG) suggests that not a strong surface field is 
required if the global structure of the stellar magnetic field was 
according to the Parker's model \citep{parker_58}. 

We note that such not strong magnetic fields are likely present in
massive hot stars and the clear sign for this is the non-thermal radio 
emission detected from wide WR$+$O binaries \citep{doug_00}. This 
emission is identified with the colliding wind shock region in these 
objects and WR 140 and WR 147 are two classical examples of those.
Interestingly, there are {\it no signs} of non-thermal component in the 
X-ray spectra of these objects which are successfully matched by thermal 
emission from plasma behind strong adiabatic shocks
(e.g. for WR 140 see \citealt{zhsk_00}; \citealt{pollock_05};
for WR 147 see \citealt{zhek_07}; \citealt{zhp_10}). 
Thus, a physical picture where strong adiabatic shocks produce thermal
X-rays and the corresponding DSA of relativistic particles is not very 
efficient does not seem to be anything unique and it might well be a 
reasonable assumption for the current evolution stage of \snr.

Finally, the ongoing monitoring of the X-ray emission from \snr 
will be very helpful in further constraining the physical picture
described here.
First,  
the continuous brightening of \snr will allows us to establish with
higher certainty the level of non-thermal X-ray emission from this 
very young supernova remnant.
We note that if strong (and increasing) non-thermal emission is 
detected, it will require considerable changes in the physical picture
as we described it in this work.
Second, the data from the monitoring observations will
also provide us with detailed spectral information (grating data)
that will reveal the evolution of the kinematics of the X-ray emitting
plasma.
We note that the RSS picture was proposed to explain the relatively
narrow spectral lines in the grating spectra of \snrE. As seen from
Figs.~\ref{fig:flux1} and \ref{fig:flux2}, the contribution from the
reflected shock to the total observed fluxes is not dominant even at
times after 6,000~DAE which might be a deficiency from the poor 
spectral resolution in CCD spectra.
Upgrading the available CCD data base with grating spectra, having good 
quality and spanning over long enough period of time, will allow us to 
test the global RSS model more rigorously by confronting its 
predictions with the actual evolution of the gas kinematics of the
X-ray emitting plasma. 
This will also justify amendments to the model by taking into
account the X-ray emission from the region behind the reverse shock.
In turn, we can further improve our
understanding of the exciting phenomenon of the birth of supernova
remnant 1987A.

\section{Conclusions}
\label{sec:conclusions}
In this study, we continued to explore the validity of the
reflected-shock structure picture in \snrE. This picture assumes that
the blast wave is interacting with the HII region in
vicinity of the inner ring and when it encounters dense clump(s) 
a shock is transmitted into the clump(s) and a reflected shock forms
that additionally compresses the shocked gas behind the blast wave.
We adopted an improved version of the reflected-shock model (global RSS
model) to test the RSS picture in \snr by making
use of CCD spectra from the monitoring program with {\it Chandra}.
The main results from this analysis are as follows.

1. The global RSS model is capable of matching both the X-ray expansion
velocity curve \citep{judy_09} as well as the X-ray spectra of 
\snr over its evolution for the last ten years or so. Moreover,
extrapolating back in time the X-ray flux, the model is able to match 
the observed {\it ROSAT} fluxes from the time of the supernova 
reappearance in X-rays (1000-3000 DAE).

2. The evolution of the mean electron temperature behind the shocks 
shows a decrease  with time of the electron temperature
in the blast wave, while a slight increase with time is noticed  
for this parameter in the transmitted shock. The mean ratio of the
electron to the mean plasma temperature is required to increase with
the decrease of the velocity of the blast wave.

3. At early times, the soft (0.5-2 keV) X-ray flux of \snr was 
dominated by the blast wave but for times after $\sim 6000$ DAE 
the transmitted shock became dominant. This transition coincides 
with the steep upturn in the X-ray LC reported by \citet{sang_05}.
Also, the emission measure of the hot plasma in the
transmitted shock (as well as in the reflected shock) is increasing 
steeper than that for the blast wave for times after $\sim 6400$ DAE. 
All this is indicative of an ongoing penetration of the blast wave 
deeper in the equatorial ring debris.

4. Results from the X-ray analysis were used to explore
the possibility that the X-rays and the non-thermal radio emission from
\snr originate from the same shock structure (the blast wave).
Using a simplified analysis, it was shown that this can indeed be the
case, provided the particle acceleration mechanisms, that operate in
the shocks in \snrE, are not very efficient in producing relativistic
particles with very high energies.

5. As a next step, the global RSS model could be used to obtain 
further constraints on the reflected-shock structure picture in 
\snr by analyzing X-ray spectra with good spectral resolution once 
the available data cover a long enough time interval of the \snr 
evolution. We believe that this will allow us to improve our 
understanding of the underlying physics and to build a more 
realistic picture and model of this fascinating phenomenon: the birth 
and evolution of supernova remnant 1987A.

\section{Acknowledgments}
SAZ acknowledges financial support from Bulgarian National Science
Fund grant DO-02-85.
The authors thank an anonymous referee for valuable comments and
criticism.

\bsp

\label{lastpage}

\end{document}